\documentclass[aps,prl,twocolumn,showpacs,preprintnumbers,amsmath,amssymb,superscriptaddress]{revtex4}%
\usepackage{graphicx}% Include figure files
\usepackage{dcolumn}% Align table columns on decimal point
\usepackage{bm}% bold math
\usepackage{color}
\begin{document}

\preprint{preprint(\today)}

\title{Spin-lattice coupling induced weak dynamical magnetism in EuTiO$_{3}$ at high temperatures}

\author{Z.~Guguchia}
\affiliation{Physik-Institut der
Universit\"{a}t Z\"{u}rich, Winterthurerstrasse 190, CH-8057
Z\"{u}rich, Switzerland}

%\author{R.~Khasanov}
%\affiliation{Laboratory for Muon Spin Spectroscopy, Paul Scherrer Institute, CH-5232
%Villigen PSI, Switzerland}

%\author{A.~Shengelaya}
%\affiliation{Department of Physics, Tbilisi State University,
%Chavchavadze 3, GE-0128 Tbilisi, Georgia}

\author{H.~Keller}
\affiliation{Physik-Institut der Universit\"{a}t Z\"{u}rich,
Winterthurerstrasse 190, CH-8057 Z\"{u}rich, Switzerland}

\author{R.K.~Kremer}
\affiliation{Max Planck Institute for Solid State Research, Heisenbergstr.~1, D-70569 Stuttgart, Germany}

\author{J.~K\"{o}hler}
\affiliation{Max Planck Institute for Solid State Research, Heisenbergstr.~1, D-70569 Stuttgart, Germany}

\author{H.~Luetkens}
\affiliation{Laboratory for Muon Spin Spectroscopy, Paul Scherrer Institute, CH-5232
Villigen PSI, Switzerland}

\author{T.~Goko}
\affiliation{Laboratory for Muon Spin Spectroscopy, Paul Scherrer Institute, CH-5232
Villigen PSI, Switzerland}

\author{A.~Amato}
\affiliation{Laboratory for Muon Spin Spectroscopy, Paul Scherrer Institute, CH-5232
Villigen PSI, Switzerland}

\author{A.~Bussmann-Holder}
\affiliation{Max Planck Institute for Solid State Research, Heisenbergstr.~1, D-70569 Stuttgart, Germany}

\begin{abstract}
 
EuTiO$_{3}$, which is a G-type antiferromagnet below $T_{\rm N}$ = 5.5 K, has some fascinating properties at 
high temperatures, suggesting that macroscopically hidden dynamically fluctuating weak 
magnetism exists at high temperatures. This conjecture is substantiated by magnetic 
field dependent magnetization measurements, which exhibit pronounced anomalies below 200 K 
becoming more distinctive with increasing magnetic field strength. Additional results from 
muon spin rotation (${\mu}$SR) experiments provide evidence for weak fluctuating bulk magnetism induced by 
spin-lattice coupling which is strongly supported in increasing magnetic field.

\end{abstract}


\maketitle

Perovskite oxides are well known for their rich ground states and the possibility of tuning these by doping, pressure, 
and temperature. Especially, their ferroelectric properties have attracted increased interest, since they offer a broad 
range of technological applications [\cite{Scott} and Refs. therein]. EuTiO$_{3}$ (ETO) has first been synthesized in the early fifties when a major boom 
in the search of ferroelectrics without hydrogen bonds took place \cite{Brous}. Since ETO did not show any ferroelectric properties, 
but became antiferromagnetic (AFM) at low temperature with $T_{N}$ ${\approx}$ 5.5 K \cite{McGuire,Chien}, it vanished from research activities and only regained 
substantial interest lately when it was demonstrated that strong magneto-electric coupling is present in this material \cite{Katsufuji}. 
This was established by dielectric permitivity measurements, where an anomaly in the dielectric permitivity  ${\varepsilon}$ sets in at $T_{\rm N}$. 
This anomaly vanishes upon the application of a magnetic field. The rather large and strongly temperature dependent dielectric 
permitivity has been related to a soft optic q = 0 mode, reminiscent of a ferroelectric soft mode \cite{Kamba,Goian}. However, its complete freezing is
inhibited by quantum fluctuations quite analogous to SrTiO$_{3}$ (STO) \cite{Muller}. In the search for new multiferroic materials ETO 
thus became a potential candidate and enormously enhanced the research in the ETO physical and chemical properties. 
In the focus of the research was mainly the transition to the AFM phase \cite{Scagnoli,Allieta} and its possible polar properties 
with novel spin arrangements together with electronic structure investigations \cite{Bettis}.\\
%%%%%%%%%%%%%%%%%%%%%%%%%%%%%%%%%%%%%%%%%%%%%%%%%%%%%%%%%%%%%%
\begin{figure}[ht!]
\includegraphics[width=1.0\linewidth]{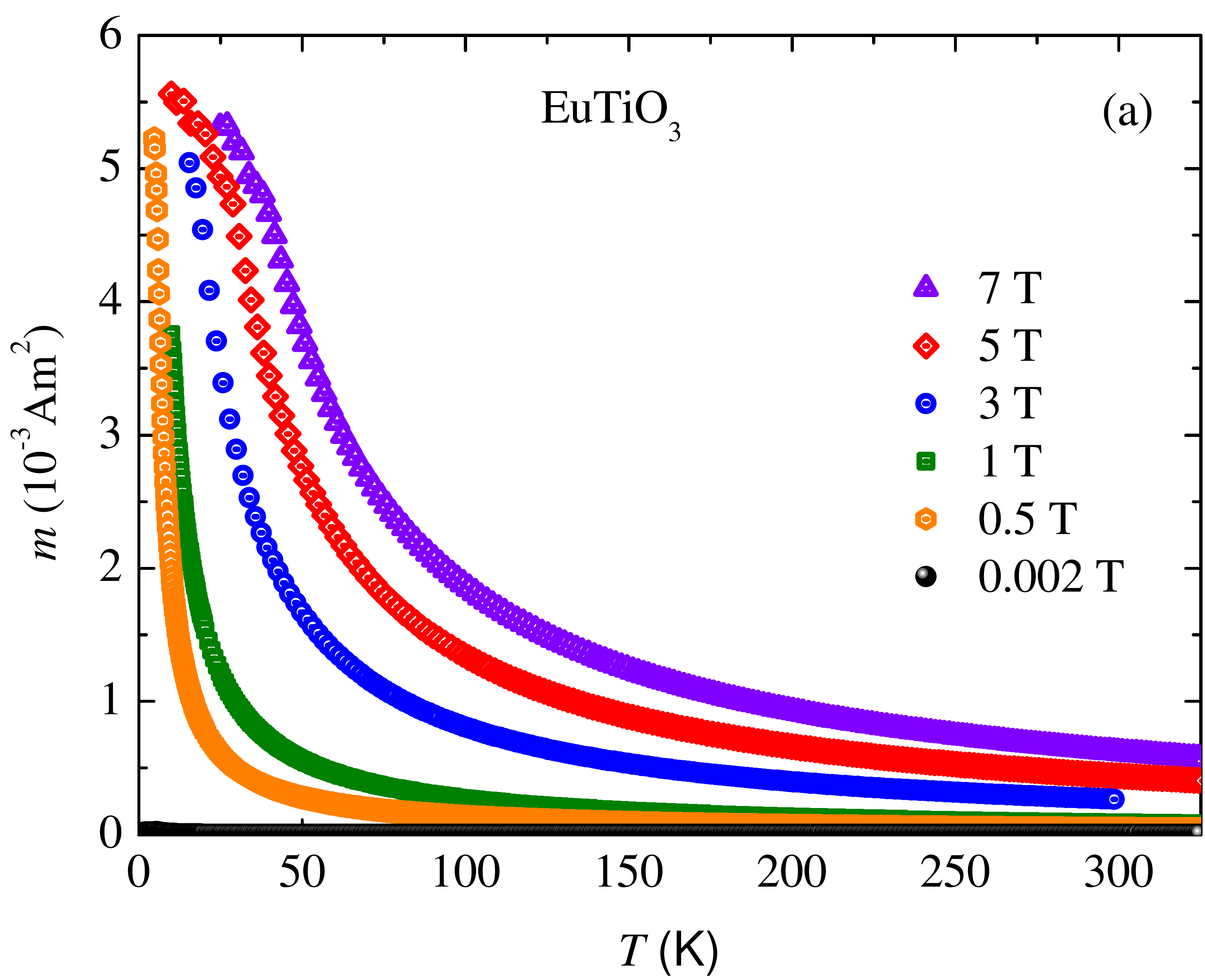}
\includegraphics[width=1.0\linewidth]{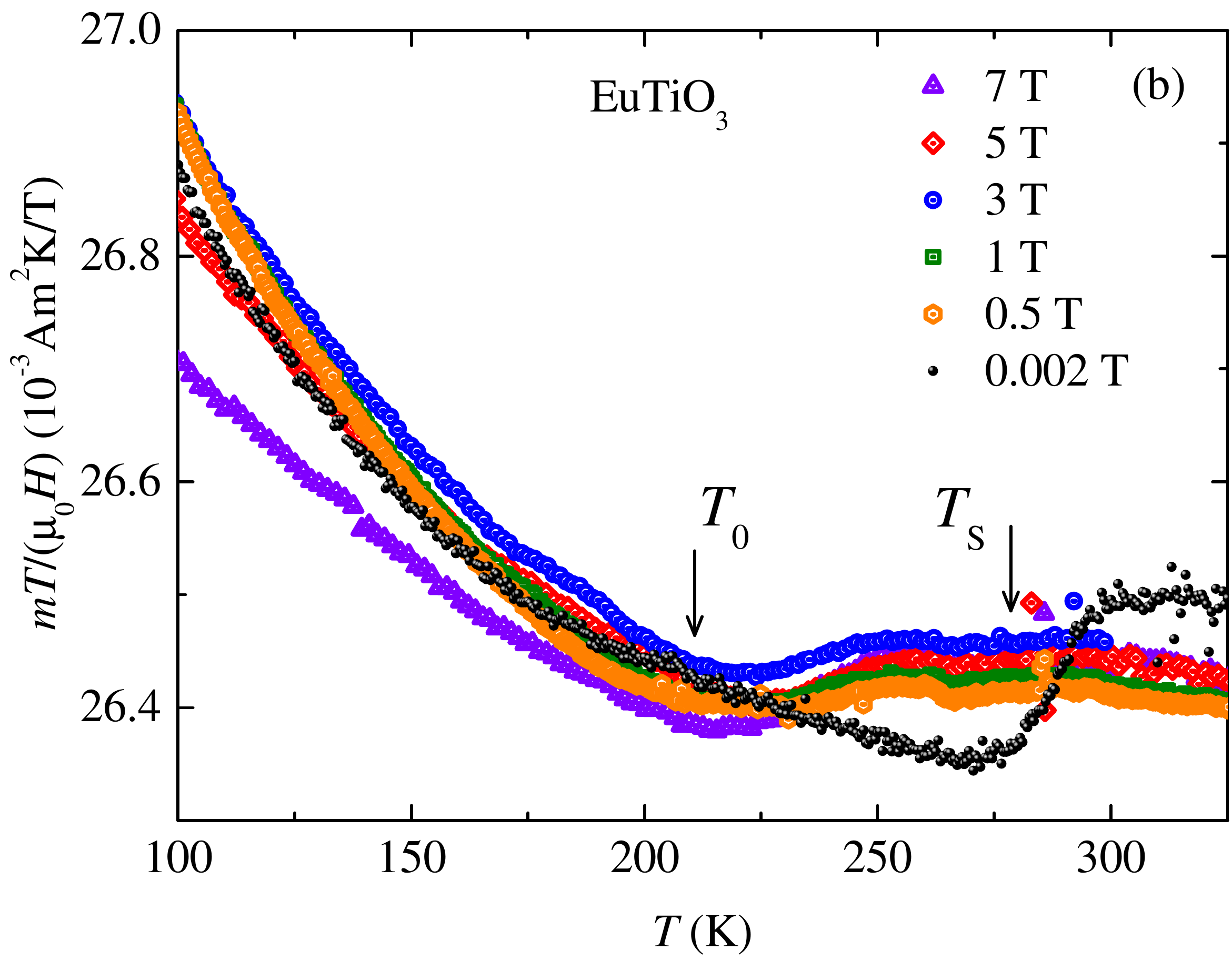}
\vspace{-0.4cm}
\caption{ (Color online) (a) The magnetic moment $m$ of EuTiO$_{3}$ as a function of temperature and magnetic field.  
(b) The quantity $mT/({\mu}_{0}H)$ as a function of temperature and magnetic field. 
The arrows indicate the structural phase transition temperature $T_{\rm s}$ and the
magnetic crossover temperature $T_{\rm 0}$.}
\label{fig1}
\end{figure}
%%%%%%%%%%%%%%%% 
%%%%%%%%%%%%%%%%%%%%%%%%%%%%%%%%%%%%%%%%%%%%%%%%%%%%%%%%%%%%%%
\begin{figure}[t!]
\centering
\includegraphics[width=1.0\linewidth]{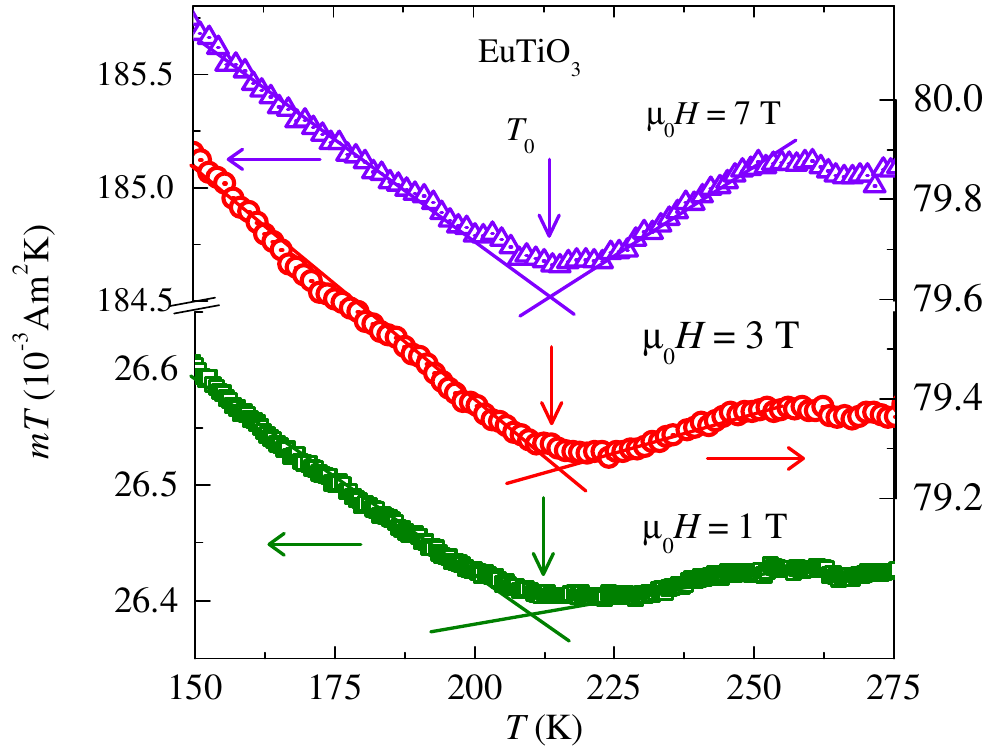}
\vspace{-0.8cm}
\caption{ (Color online) Magnetic moment $m$ multiplied by temperature $T$ of EuTiO$_{3}$ for ${\mu}_{0}$$H$ = 1, 3, and 7 T.
The arrows indicate the magnetic crossover temperature $T_{\rm 0}$.}
\label{fig1}
\end{figure}
%%%%%%%%%%%%%%%% 
The close analogy between ETO and STO, namely the same lattice constants, the same ionic valency of the 
cation with almost equal ionic radii, and the tendency towards a polar instability, has recently been shown to 
be even closer, by establishing that ETO transforms from cubic to tetragonal caused by an oxygen octahedral 
rotation instability \cite{Bussmann-Holder1}. Amazingly, this takes place at $T_{\rm S}$ = 282 K in ETO, in contrast to 
STO with $T_{\rm S}$ = 105 K. 
This large spread in $T_{\rm S}$ between the two compounds can only be caused by the different atomic masses of Sr and Eu and 
also by the Eu 4$f$ electrons with spin $S$ = 7/2. In order to obtain a more clear picture of the 
origin of this difference in $T_{\rm S}$, the mixed crystal series Sr$_{x}$Eu$_{1-x}$TiO$_{3}$ has been studied as a 
function of $x$ with focus on the development of $T_{\rm S}$ and $T_{\rm N}$ with $x$ \cite{Bussmann-Holder2,Guguchia2,Guguchia3}. 
Interestingly, both transition temperatures vary nonlinearly with $x$, reflecting the dilution effect of the Eu 4$f$ spins. 
Especially, the low temperature phase transition line 
indicates the stability of the AFM order which persists up to $x$ ${\sim}$ 0.25, in contrast to recent 
theoretical arguments which predict a transition from AFM via ferrimagnetic to ferromagnetic 
order with decreasing $x$ \cite{Morozovska,Eliseev2}. The high temperature phase transition line has been obtained by different 
experimental techniques, namely, electron paramagnetic resonance (EPR), muon spin rotation (${\mu}$SR), electrical resistivity, and 
specific heat measurements \cite{Bussmann-Holder2}. The spectacular aspect stems from EPR and ${\mu}$SR data which test magnetic properties. 
Especially, a finite ${\mu}$SR relaxation rate signals the presence of some kind of exotic magnetism being present in the bulk 
sample, and indicates that dynamic magnetic order must be present in spatially confined regions of 
the ceramics. This conclusion is further supported by the fact that $T_{\rm S}$  depends on an applied external 
magnetic field which is amazing, since $T_{\rm S}$ lies 275 K above $T_{\rm N}$ \cite{Guguchia24}, $i.e.$, deeply in the paramagnetic region. 
From both results it must be concluded that the oxygen octahedral rotations influence the Eu 4$f$ spin-spin interactions still at high 
temperatures and contribute to an effective second nearest neighbor ferromagnetic spin exchange which is otherwise small or vanishing.\\
%%%%%%%%%%%%%%%%%%%%%%%%%%%%%%%%%%%%%%%%%%%%%%%%%%%%%%%%%%%%%%
\begin{figure}[b!]
\includegraphics[width=1.0\linewidth]{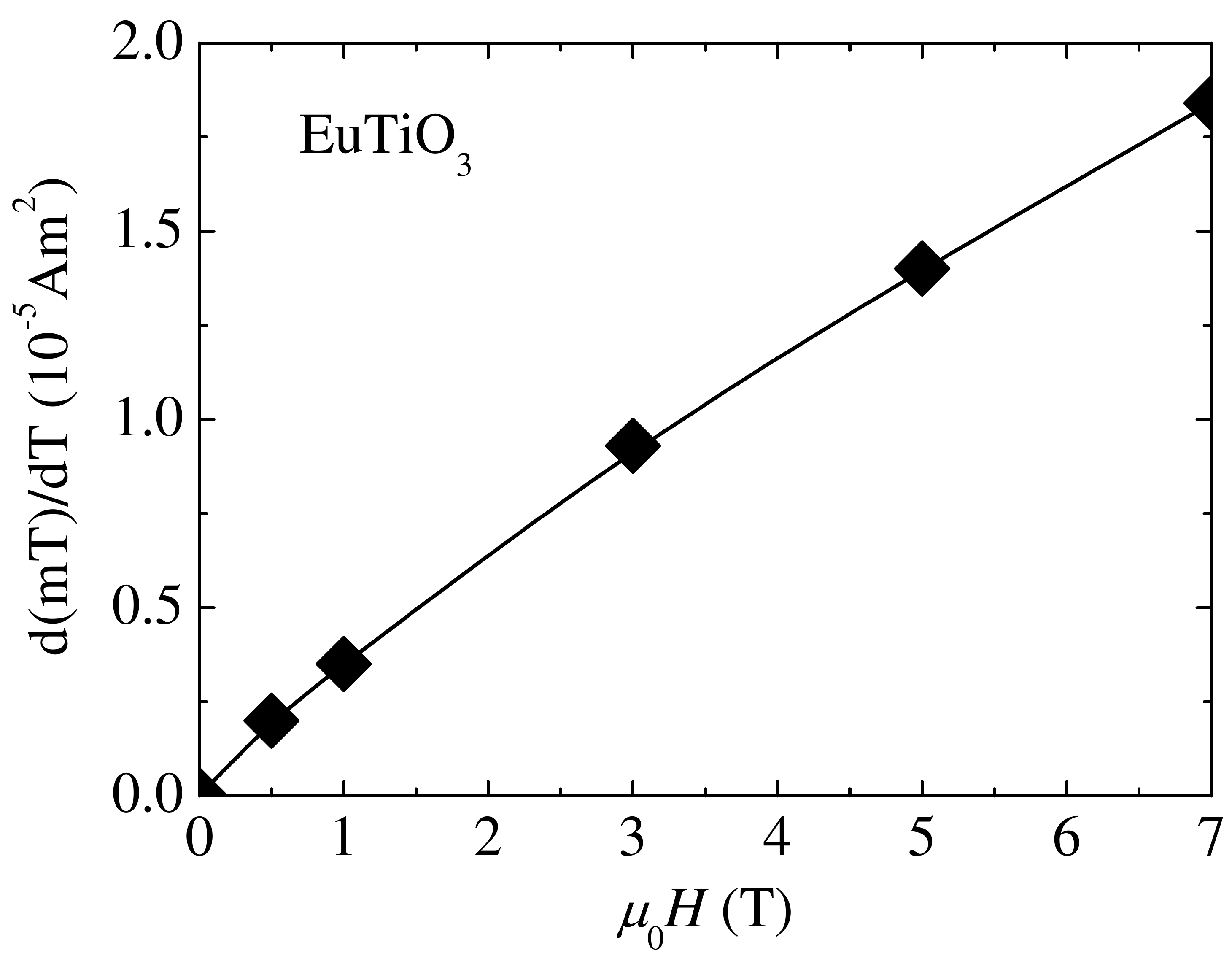}
\vspace{-0.8cm}
\caption{ (Color online) The slope of the magnetic moment times temperature data $mT$ (Figs. 2) 
as a function of the applied magnetic field ${\mu}_{0}H$ for EuTiO$_{3}$ as obtained in the temperature range between 150 and 200 K. The full line in the figure is a guide to the eye. The size of the symbols exceeds the experimental error bars.}
\label{fig1}
\end{figure}
%%%%%%%%%%%%%%%% 
To substantiate these conclusions, we have recently proposed a novel approach to test strong spin-lattice 
couplings by magnetization measurements \cite{Caslin}. If a coupling between the spins and the lattice 
is present in the form of a quadratic interaction $E_{SL}$ ${\sim}$ ${\alpha}$${\sum_{n,i,j}}$w$^{2}$$S_{i}$$S_{j}$, where  w  
is the polarizability coordinate of the TiO$_{6}$ cluster and $S_{i}$,$S_{j}$   
are second nearest neighbor spins along the diagonal via the intermediate oxygen ions, with ${\alpha}$ being the coupling 
strength, then the magnetic susceptibility attains an extra temperature dependent component in the paramagnetic phase according to: 
${\chi}$ = $N$$\mu_{0}$$g^{2}$${\mu}_{B}^{2}$S(S+1)(1+${\alpha}$$\langle$w$\rangle$$_{T}^{2}$)/(3$k_{\rm B}$$T$) 
with $N$, ${\mu}_{\rm B}$, $g$ being the number of magnetic moments per volume, the Bohr magneton and the $g$ factor, respectively.
The polarizability coordinate refers to the relative displacement coordinate between core and shell of the nonlinearly polarizable 
cluster mass at the TiO$_{3}$ lattice site and its role in phase transitions has been discussed in detail in Refs. \cite{Bussmann-Holdernew, Migoni, Bilzh}. 
Since $\langle$w$\rangle$$_{T}^{2}$ is the thermal average over the polarizability coordinate into which all dynamical information enters, 
the temperature dependence of the soft transverse acoustic zone boundary mode contributes essentially to its temperature dependence. 
Above $T_{\rm S}$ the squared soft mode frequency decreases 
linearly with decreasing temperature to become zero at $T_{\rm S}$. Below $T_{\rm S}$ the mode recovers to increase in a 
Curie-Weiss type manner linearly with twice the gradient as compared to the para phase. 
This scenario has the consequence that the product of susceptibility and temperature is not a constant far above
$T_{\rm N}$ but follows the $T$-dependence of the soft mode which we indeed were able to demonstrate recently \cite{Guguchia24}. 
However, in addition another consequence results, since the increasing size of $\langle$w$\rangle$$_{T}^{2}$ 
below $T_{S}$, respectively the increasing rotation angle, leads to an increasing strength in the ferromagnetic 
second nearest neighbor exchange. This supports the formation of growing dynamical magnetic clusters which 
can be tested by applying a magnetic field. Indirectly cluster formation has already been seen in the bare magnetic susceptibility 
data where an upturn was detected approximately 100 K below $T_{S}$ \cite{Guguchia24}.
 %%%%%%%%%%%%%%%%%%%%%%%%%%%%%%%%%%%%%%%%%%%%%%%%%%%%%%%%%%%%%%
\begin{figure}[b!]
\includegraphics[width=1.0\linewidth]{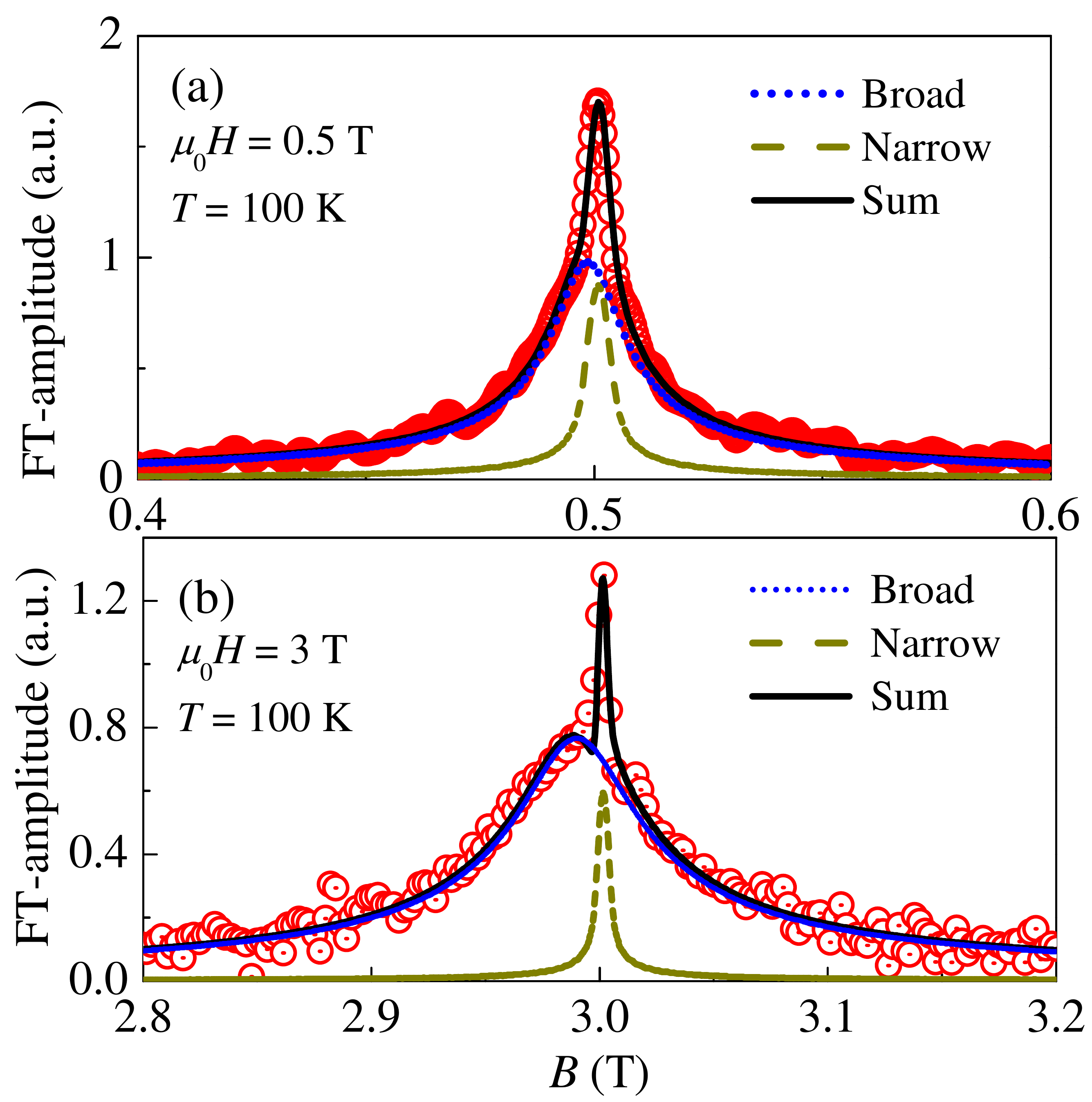}
\vspace{-0.8cm}
\caption{ (Color online) Fourier transform for the ${\mu}$SR asymmetry spectra of EuTiO$_{3}$ at 100 K for two magnetic fields: 
(a) ${\mu}_{0}$$H$ = 0.5 T and (b) 3 T. The solid lines are the FTs of the corresponding theoretical $A(t)$ given by Eq.~(1).}
\label{fig1}
\end{figure}
%%%%%%%%%%%%%%%% 
In order to substantiate this scenario, 
we have performed magnetization measurements in applied magnetic fields up to ${\mu}_{0}$$H$ = 7 Tesla
on ceramic samples of ETO which have been prepared as described in Ref. \cite{Bussmann-Holder1}. 
The magnetic moment $m$ as a function of temperature and at various magnetic fields is shown
in Fig.~1a. While the data in Fig.~1a appear to be typical for a paramagnetic system, 
a detailed analysis of the data in terms of the product $m$$T$ normalized by the magnetic field ${\mu}_{0}$$H$ 
versus $T$ reveals striking anomalies which - for the smallest field ${\mu}_{0}$$H$ = 0.002 T - are related to the structural 
instability as already reported before. With increasing field strength a second temperature scale emerges from the 
data appearing around $T_{\rm 0}$ ${\simeq}$ 210 K which does not depend on the field strength (Fig.~1b). In the following we denote $T_{\rm 0}$ as the magnetic crossover temperature.
The normalization of the data with respect to $H$ has the advantage that data for all field strengths 
can be shown in a single graph. However, this methodology obscures important details which are striking when  
$m$$T$ is plotted as a function of temperature. The details are shown in Fig.~2 for three representative field strengths. 
Apparently, in Fig.~2 a change in slope (in the temperature range between 150 and 210 K)
of $m$$T$ versus $T$ takes place with increasing field 
strength which is shown in detail and for all field strengths in Fig.~3. With increasing field 
strength the slope increases nonlinearly with the magnetic field, evidencing that some kind of correlated magnetism is present. 
Since the oxygen octahedral rotations modify the second nearest neighbor ferromagnetic exchange interaction only 
and have no influence on the direct nearest neighbor AFM exchange, we suggest that weak ferromagnetism appears below 
~210 K. Note, that this finding has nothing to do with structural modulations as reported in Refs. \cite{Allieta} and \cite{KimJW}, since the x-ray tested temperature dependent structural data are only in accordance with tetragonal symmetry and show no superlattice reflections. It is also important to note that the data in \cite{Allieta} report a $T_{\rm S}$ around 245 K, much lower than our value, which has been corrected
 in a subsequent publication to arrive at the same value as ours. In Ref. \cite{KimJW} single crystals have been investigated with $T_{\rm N}$
 being substantially smaller than our and other literature values, which might stem from impurities or defects and thus can give rise 
to the reported structural modulations. In addition, a recent detailed structural study confirmed our results and arrived at the conclusion
that the unmodulated case corresponds to the bulk structure of pure material \cite{Ellisnew}.  The weak magnetism
has been tested by bulk sensitive local probe experiments as provided by the ${\mu}$SR technique. 
As has already been demonstrated in Ref.~12, a finite ${\mu}$SR relaxation rate is seen above 
the structural phase transition temperature and persists in the tetragonal phase.\\

 These former data are here complemented by measurements of the ${\mu}$SR relaxation rate as a function of 
temperature and magnetic field. Transverse-field (TF) ${\mu}$SR 
experiments were performed at the ${\pi}$E3 beamline of the Paul Scherrer
Institute (Villigen, Switzerland), using the HAL-9500 ${\mu}$SR spectrometer.
The specimen was mounted in a He gas-flow cryostat with the largest
face perpendicular to the muon beam direction, along which the external field was applied.
Magnetic fields between 10 mT and 5 T were applied, and the temperatures were varied between
100 and 300 K. The ${\mu}$SR time spectra were analyzed using the free software package MUSRFIT \cite{Suter}.\\ 
%Longitudinal-field (LF) ${\mu}$SR experiments were carried out at the ${\pi}$M3 beamline
%using the general purpose instrument (GPS).\\
%%%%%%%%%%%%%%%%%%%%%%%%%%%%%%%%%%%%%%%%%%%%%%%%%%%%%%%%%%%%%%
\begin{figure}[t!]
\includegraphics[width=1.0\linewidth]{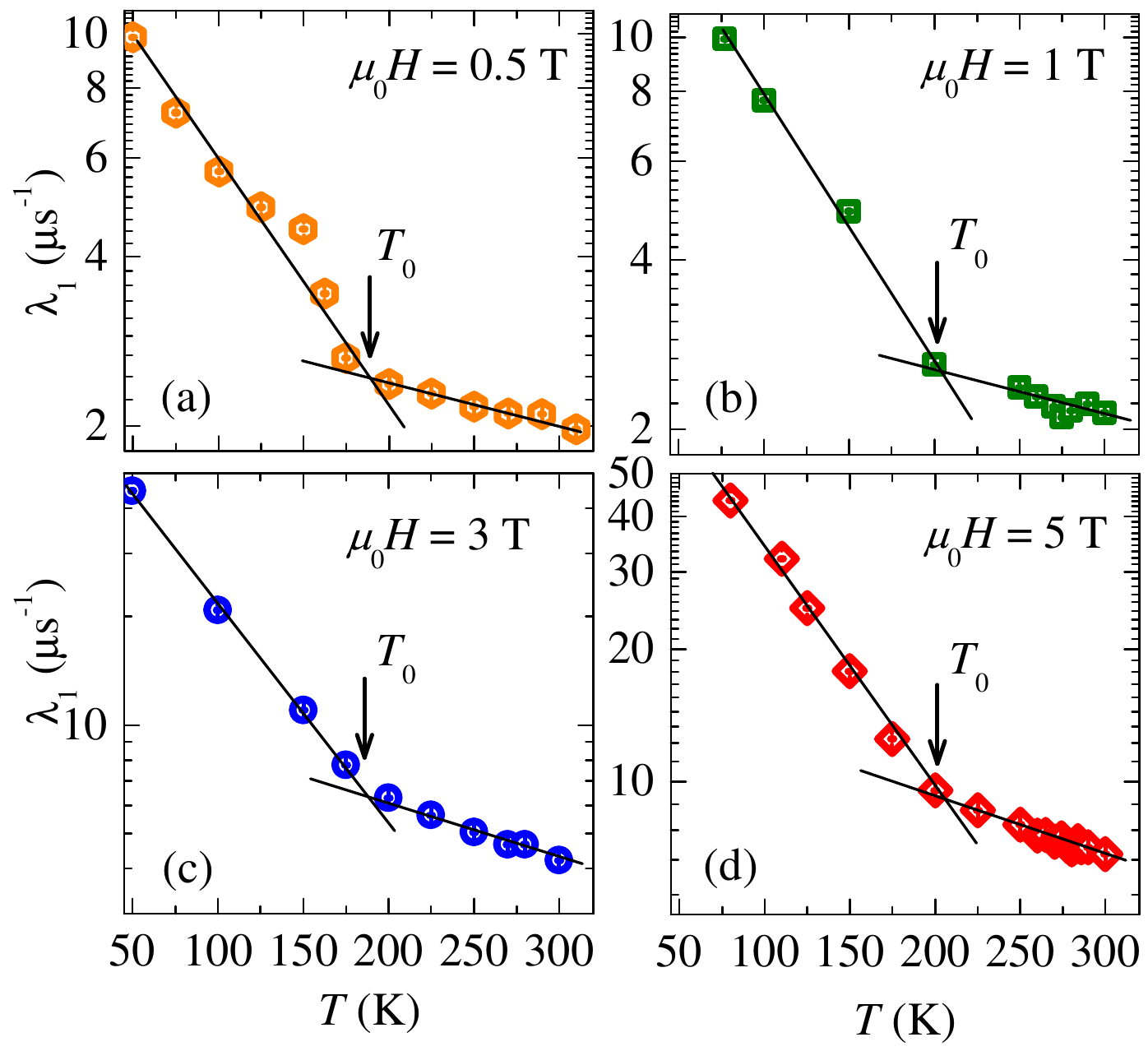}
\vspace{-0.8cm}
\caption{ (Color online) ${\mu}$SR relaxation rate ${\lambda}_{1}$ of EuTiO$_{3}$ as a function of temperature measured for various magnetic fields:
(a) ${\mu}_{0}H$ = 0.5 T, (b) ${\mu}_{0}H$ = 1 T, (c) ${\mu}_{0}H$ = 3 T, and (d) ${\mu}_{0}H$ = 5 T.
The thin lines  are guides to the eyes. The arrow indicates the magnetic crossover temperature $T_{0}$ where the kink in ${\lambda}_{1}$ occurs. The experimental error bars are smaller than the size of the symbols.}
\label{fig1}
\end{figure}
%%%%%%%%%%%%%%%% 
%%%%%%%%%%%%%%%%%%%%%%%%%%%%%%%%%%%%%%%%%%%%%%%%%%%%%%%%%%%%%%
\begin{figure}[b!]
\includegraphics[width=1.0\linewidth]{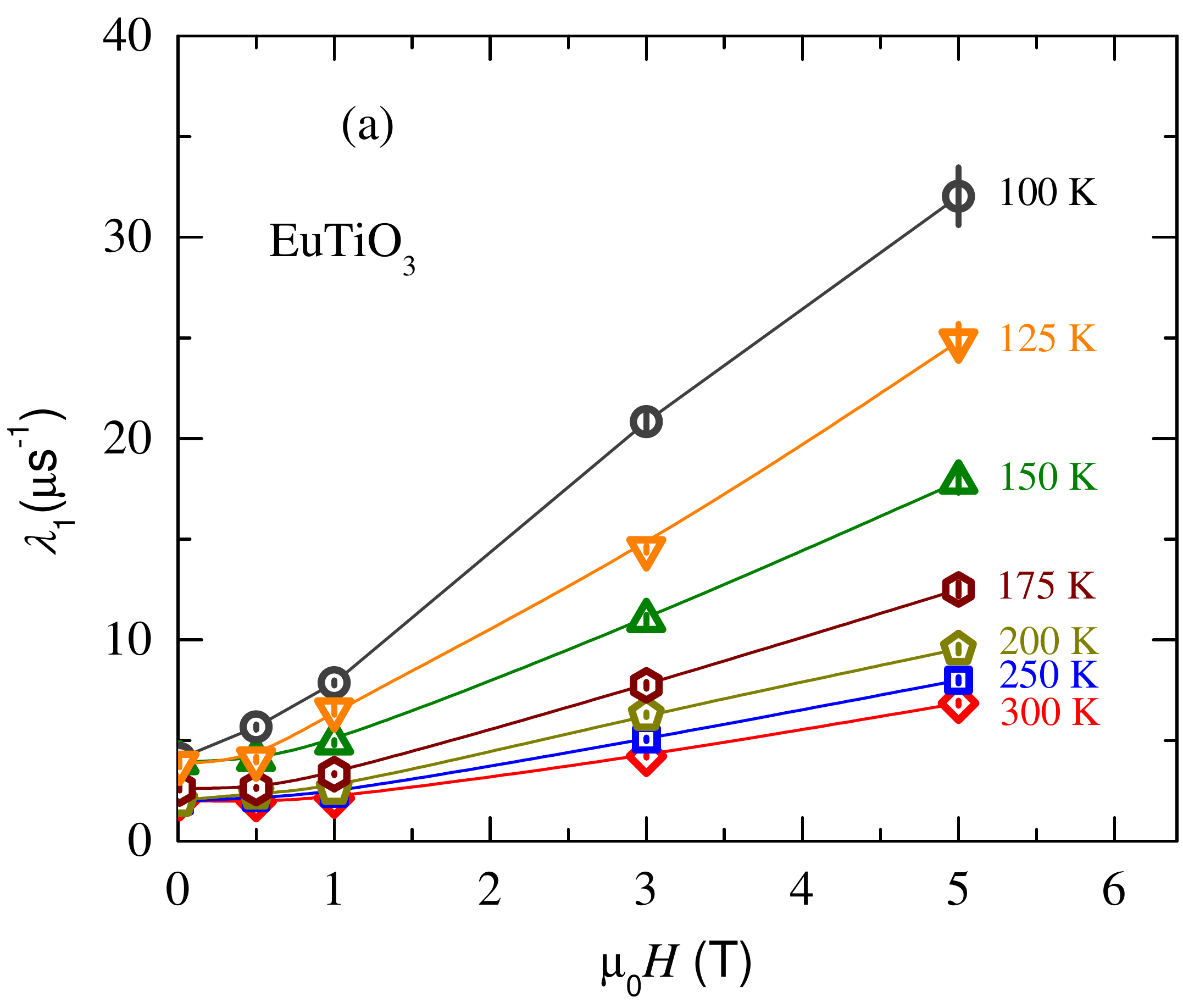}
\includegraphics[width=1.0\linewidth]{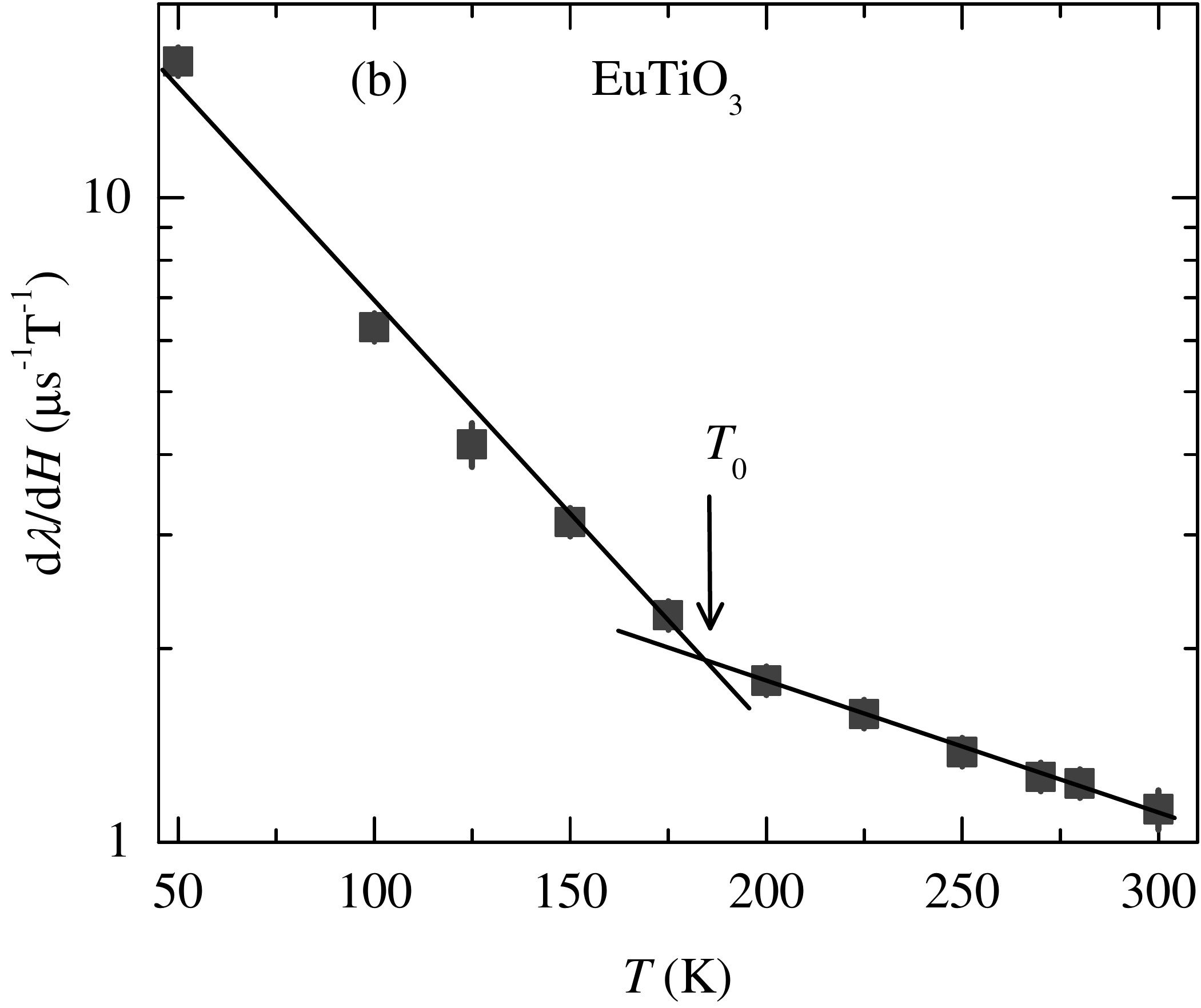}
\vspace{-0.8cm}
\caption{ (Color online) (a) ${\mu}$SR relaxation rate ${\lambda}_{1}$ of EuTiO$_{3}$ as a function of magnetic field for various temperatures. The solid lines are guides to the eyes. (b) The temperature dependence of the 
quantity d${\lambda}$/d$H$. The arrow indicates the magnetic crossover temperature $T_{0}$.}
\label{fig1}
\end{figure}
%%%%%%%%%%%%%%%% 
 For all fields and in the whole investigated temperature range, two-component signals were observed: 
a signal with fast exponential relaxation (broad signal) and another one with a slow exponential relaxation (narrow one). 
As an example, the Fourier transform (FT) of the ${\mu}$SR asymmetry at 100 K and for 0.5 T and 3 T is shown in Fig.~4. 
The ${\mu}$SR time spectra were analyzed by using the following functional form:
%%%%%%%%%%%%%%%%%%%%%%%%%%%%%%%%%%%%%%%%%%%%%%%%%%%%%%%%%%%%%%
\begin{equation}
\begin{split}
A(t)=A_1\exp(-\lambda_1t)\cos(\gamma_{\mu}B_{\mu1}t+\varphi)+ \\
A_2\exp(-\lambda_2t)\cos(\gamma_{\mu}B_{\mu2}t+\varphi), 
\end{split}
\label{eq1}
\end{equation}
%%%%%%%%%%%%%%%%%%%%%%%%%%%%%%%%%%%%%%%%%%%%%%%%%%%%%%%%%%%
where $A_{\rm 1}$($A_{\rm 2}$), $B_{\rm \mu1}$($B_{\rm \mu2}$), and ${\lambda}_{1}$(${\lambda}_{2}$)
denote the assymmetry, the local magnetic field at muon site, and the relaxation rate of the fast (slow) component.
%The relaxation rates ${\lambda}_{1}$ and ${\lambda}_{2}$ characterize the exponential damping of the two components.
$\gamma/(2{\pi})\simeq 135.5$~MHz/T is the muon gyromagnetic ratio, and ${\varphi}$ is the initial phase of the muon-spin ensemble. 
Regarding the two-component signals of EuTiO$_{3}$, the signal with the fast (slow) relaxation is associated with the
volume fraction with (without or only very weak) magnetic order. Since the major fraction (${\simeq}$ 80 ${\%}$) of the ${\mu}$SR signal comes from the
muons stopping in the part of the sample with fast relaxation, 
we discuss here only the temperature and field dependence of the 
relaxation rate ${\lambda}_{1}$ of the fast (magnetic) component.

%%%%%%%%%%%%%%%%%%%%%%%%%%%%%%%%%%%%%%%%%%%%%%%%%%%%%%%%%%%%%%
%\begin{figure}[t!]
%\includegraphics[width=1.0\linewidth]{ETO-LF.pdf}
%\vspace{-0.3cm}
%\caption{ (Color online) Longitudinal field dependence of muon-spin relaxation in EuTiO$_{3}$ at 130 K. The inset shows the raw data at two fields.}
%\label{fig1}
%\end{figure}
%%%%%%%%%%%%%%%% 
%The fast depolarization of the ${\mu}$SR signal observed in  EuTiO$_{3}$ could be either due to a wide distribution
%of static fields, and/or to strongly fluctuating magnetic
%moments. To discriminate between these two, ${\mu}$SR experiments of EuTiO$_{3}$ under longitudinal field (LF)
%up to 0.5 T were performed at temperatures below $T_{0}$. 
%Since no recovery of the muon polarization is observed at long times we conclude that in EuTiO$_{3}$ 
%dynamically fast fluctuating ferromagnetism is present. 

The temperature dependence of  ${\lambda}_{\rm 1}$ for various applied fields is shown in Fig.~5.
It is evident that ${\lambda}_{1}$($T$) exhibits a pronounced kink at 
the magnetic crossover temperature $T_{\rm 0}$ ${\simeq}$ 200 K for all measured fields.
Note that this value of $T_{\rm 0}$ is smaller than $T_{\rm 0}$ ${\simeq}$ 210 K derived from magnetization
measurements (see Fig. 2). This is likely due to the different experimental techniques (magnetization and ${\mu}$SR) used to
determine $T_{\rm 0}$.
Between 300 and 200 K ${\lambda}_{1}$($T$) increases with decreasing temperature.
For $T > T_{0}$, an additional significant increase of ${\lambda}_{1}$($T$) is observed.
Moreover, a strong enhancement of ${\lambda}_{1}$ with increasing field was found at all investigated temperatures (see Fig.~6a). 
At low temperatures ${\lambda}_{\rm 1}$($H$) increases stronger than at high temperatures.
For clarity, the quantity d${\lambda}$/d$H$ determined from the linear part of ${\lambda}$($H$) is plotted in Fig.~6b 
as a function of temperature. A change in the slope of the temperature dependence of d${\lambda}$/d$H$ can be clearly seen at 
$T_{\rm 0}$ ${\simeq}$ 200 K. The pronounced increase 
and the stronger field dependence of ${\lambda}_{\rm 1}$  below $T_{\rm 0}$ suggests the 
appearance of some kind of magnetic correlations in the system. The formation of field induced magnetic clusters below $T_{\rm 0}$
may be a possible explanation as proposed in Ref.~\cite{Caslin}.     
Unfortunately, our data do not admit to draw any definite conclusions on the type of magnetic correlations (AFM or FM).
However, we can figure out whether the fast depolarization of the ${\mu}$SR signal observed
below $T_{\rm 0}$ in  EuTiO$_{3}$ is either due to a broad distribution of static fields, and/or to strongly fluctuating magnetic
moments. In order to discriminate between these two cases ${\mu}$SR experiments in longitudinal fields (LF) \cite{Yaouanc}
are required. Therefore, we performed LF ${\mu}$SR experiments on ETO in fields up to 0.5 T below $T_{0}$. 
Since no recovery of the muon polarization was observed at long times we conclude that in EuTiO$_{3}$ 
dynamic magnetism is present. 

To conclude, magnetization measurements and ${\mu}$SR data as functions of temperature 
and magnetic field provide evidence for dynamic weak magnetic clusters forming below $T_{\rm 0}$ ${\simeq}$ 200 K. 
While from magnetization an anomalous upturn in the product $m$$T$ appears around $T_{\rm 0}$ 
in a magnetic field, ${\mu}$SR data directly prove magnetic correlations which are field dependent. 
Our finding has important consequences for 
possible spintronic applications of EuTiO$_{3}$, since the magnetism far above $T_{\rm N}$ should considerably 
influence transport properties and also cause magnetic field induced changes or anomalies in dielectric constant at high temperatures 
which - until now - has not been measured. 
  
%The ${\mu}$SR experiments were performed at the Swiss Muon Source, Paul Scherrer Institute (PSI),
%Villigen, Switzerland. 
This work was supported by the Swiss National Science Foundation. 
We thank R.~Scheuermann for the support in ${\mu}$SR experiments and for valuable discussions.

%the NCCR MaNEP, and 
%the SCOPES grant No. IZ73Z0-128242, and the Georgian National Science Foundation grant RNSF/AR/10-16.

%Transition between magnetic and superconducting ground states proceeds in inhomogeneous manner. 


\begin{thebibliography}{22}

\bibitem{Scott} J. F. Scott, Science \textbf{315}, 954 (2007).

\bibitem{Brous} J. Brous, I. Frankuchen, and E. Banks, Acta Cryst. \textbf{6}, 67 (1953). 

\bibitem{McGuire} T.R. McGuire, M.W. Shafer, R.J. Joenk, H.A. Halperin, and S.J. Pickart, J. Appl. Phys. \textbf{37}, 981 (1966). 

\bibitem{Chien} L. Chien, S. DeBenedetti, and F.De.S. Barros, Phys. Rev. B \textbf{10}, 3913 (1974). 

\bibitem{Katsufuji} T. Katsufuji and H. Takagi, Phys. Rev. B \textbf{64}, 054415 (2001). 

\bibitem{Kamba} S. Kamba, D. Nuzhnyy, P. Van$\check{e}$k, M. Savinov, K. Kni$\check{z}$ek, Z. Shen, E. $\check{S}$antav$\acute{a}$, 
K. Maca, M. Sadowski, and J. Petzelt, Europhys. Lett. \textbf{80}, 27002 (2007). 

\bibitem{Goian} V. Goian, S. Kamba, J. Hlinka, P. Van$\check{e}$k, A.A. Belik, T. Kolodiazhnyi, and J. Petzelt, Eur. Phys. J. B \textbf{71}, 429 (2009). 

\bibitem{Muller} K.A. M\"{u}ller and H. Burkard, Phys. Rev. B \textbf{19}, 3593 (1979). 

\bibitem{Scagnoli} V.~Scagnoli, M. Allieta, H. Walker, M. Scavini, T. Katsufuji, L. Sagarna, O. Zaharko, and C. Mazzoli, 
Phys. Rev. B \textbf{86}, 094432 (2012). 

\bibitem{Allieta} M. Allieta, M. Scavini, L.J. Spalek, V. Scagnoli, H.C. Walker, 
C. Panagopoulos, S.S. Saxena, T. Katsufuji, and Claudio Mazzoli, Phys. Rev. B \textbf{85}, 184107 (2012).

\bibitem{Bettis} J.L. Bettis, M.-H. Whangbo, J. K\"{o}hler, A. Bussmann-Holder, and A.R. Bishop, Phys. Rev. B \textbf{84}, 184114 (2011). 

\bibitem{Bussmann-Holder1} A. Bussmann-Holder, J. K\"{o}hler, R.K. Kremer, and J.M. Law, Phys. Rev. B \textbf{83}, 212102 (2011). 

\bibitem{Bussmann-Holder2} A. Bussmann-Holder, Z. Guguchia, J. K\"{o}hler, H. Keller, A. Shengelaya, and A.R. Bishop, New Journal of Physics \textbf{14}, 093013 (2012). 

\bibitem{Guguchia2} Z. Guguchia, A. Shengelaya, H. Keller, J. K\"{o}hler, and A. Bussmann-Holder, Phys. Rev. B \textbf{85}, 134113 2012.

\bibitem{Guguchia3} Z. Guguchia, H. Keller, A. Bussmann-Holder, J. K\"{o}hler, and R.K. Kremer, 
European Physical Journal B \textbf{86}, 409 (2013).

\bibitem{Morozovska} A.N. Morozovska, Y. Gu, V.V. Khist, M.D. Glinchuk, L.-Q. Chen, V. Gopalan, and E.A. Eliseev, Phys. Rev. B \textbf{87}, 134102 (2013). 

\bibitem{Eliseev2} E.A. Eliseev, M.D. Glinchuk, V.V. Khist, C.-W. Lee, C.S. Deo, R.K. Behera, and A.N. Morozovska, J. Appl. Phys. \textbf{113}, 024107 (2013).  

\bibitem{Guguchia24} Z. Guguchia, H. Keller, J. K\"{o}hler, and A. Bussmann-Holder, J. Phys.: Condens. Matter \textbf{24}, 492201 (2012).

\bibitem{Caslin} K. Caslin, R.K. Kremer, Z. Guguchia, H. Keller, J. K\"{o}hler, and A. Bussmann-Holder, J. Phys.: Condens. Matter \textbf{26}, 022202 (2014).

\bibitem{Bussmann-Holdernew} A. Bussmann-Holder, J. Phys.: Cond. Mat. \textbf{24}, 273202 (2012).

\bibitem{Migoni} R. Migoni, H. Bilz, and D. B\"{a}uerle, Phys. Rev. Lett. \textbf{37}, 1155 (1976).

\bibitem{Bilzh} H. Bilz, G. Benedek, and A. Bussmann-Holder,  Phys. Rev. B \textbf{35}, 4840 (1987).

\bibitem{KimJW} J.-W. Kim, P. Thompson, S. Brown, P.S. Normile, J.A. Schlueter, A. Shkabko, A. Weidenkaff, and P.J. Ryan,  Phys. Rev. Lett. \textbf{110}, 027201 (2013).

\bibitem{Ellisnew} D.S. Ellis, H. Uchiyama, S. Tsutsui, K. Sugimoto, K. Kato, and A.Q.R. Baron, Physica B \textbf{442}, 34 (2014).

\bibitem{Suter} A. Suter and B.M. Wojek , $Physics~Procedia$ \textbf{30} 69-73, 2012.

\bibitem{Yaouanc} A. Yaouanc and  P. Dalmas de R\'{e}otier, J. Phys.: Cond. Mat. \textbf{9}, 9113 (1997).

%\textcolor{red}{\bibitem{Blundell} S.J. Blundell, F.L. Pratt, T. Lancaster, I.M. Marshall, C.A. Steer, S.L. Heath, J. Letard, T. Sugano, D. Mihailovic, and A. Omerzu,
% Polyhedron \textbf{22}, 1973-1980 (2003).}

%\bibitem{Dalmas1} P. Dalmas de R\'{e}otier, A. Yaouanc, P.C.M. Gubbens, S. Sakarya, E. Jiminez, P. Bonville, and J.A. Hodges, 
%Hyperfine Interactions \textbf{158}, 131 (2004).

%\bibitem{Dalmas2} P. Dalmas de R\'{e}otier, A. Yaouanc, P.C.M. Gubbens, C.T. Kaiser, C. Baines, and P.J.C. King, Phys. Rev. Lett. \textbf{91}, 167201 (2003).

%\bibitem{Orbach} R. Orbach, Proc. Roy. Soc. Lond. Sect. A \textbf{264}, 458 (1961).


\end{thebibliography}
\end{document}